# MODELLING THE INTRUSIVE FEELINGS OF ADVANCED DRIVER ASSISTANCE SYSTEMS BASED ON VEHICLE ACTIVITY LOG DATA: A CASE STUDY FOR THE LANE KEEPING ASSISTANCE SYSTEM

Kyudong Park[1)], Jiyoung Kwahk[2)*], Sung H. Han[2)], Minseok Song[2)], Dong Gu Choi[2)],
Hyeji Jang[2)], Dohyeon, Kim[2)], Young Deok Won[3)], and In Sub Jeong[3)]

[1)]Department of Creative IT Engineering, Pohang University of Science and Technology (POSTECH),
77 Cheongam-ro, Nam-gu, Pohang 37673, Korea

[2)]Department of Industrial and Management Engineering, Pohang University of Science and Technology
(POSTECH), 77 Cheongam-ro, Nam-gu, Pohang 37673, Korea

[3)]ADAS Performance Development Team, Hyundai Motor Company, 150 Hyundaiyeonguso-ro, Namyang-eup,
Hwaseong 18280, Korea

**ABSTRACT–**Although the automotive industry has been among the sectors that best-understands the importance of drivers' affect, the focus of design and research in the automotive field has long emphasized the visceral aspects of exterior and interior design. With the adoption of Advanced Driver Assistance Systems (ADAS), endowing 'semi-autonomy' to the vehicles, however, the scope of affective design should be expanded to include the behavioural aspects of the vehicle. In such a 'shared-control' system wherein the vehicle can intervene in the human driver's operations, a certain degree of 'intrusive feelings' are unavoidable. For example, when the Lane Keeping Assistance System (LKAS), one of the most popular examples of ADAS, operates the steering wheel in a dangerous situation, the driver may feel interrupted or surprised because of the abrupt torque generated by LKAS. This kind of unpleasant experience can lead to prolonged negative feelings such as irritation, anxiety, and distrust of the system. Therefore, there are increasing needs of investigating the driver's affective responses towards the vehicle's dynamic behaviour. In this study, four types of intrusive feelings caused by LKAS were identified to be proposed as a quantitative performance indicator in designing the affectively satisfactory behaviour of LKAS. A metric as well as a statistical data analysis method to quantitatively measure the intrusive feelings through the vehicle sensor log data.

**KEYWORDS** : ADAS, LKAS, Vehicle activity log data, Intrusive feeling, Affective design

## 1. INTRODUCTION

The automotive industry has long been among the sectors that best-understands the importance of drivers' affect and is making much effort to incorporate the affective attributes in their product (Schütte and Eklund, 2005; Chang *et al.*, 2006; Khalid *et al.*, 2012; Helander *et al.*, 2013). To date, affective design studies in the automotive sector have emphasized 'appearance' or 'visceral' aspects such as colour, texture, and shape of exterior (Suzianti *et al.*, 2016) and interior (Leder and Carbon 2005; Ersal *et al.*, 2011; Gkouskos and Fang, 2012) design. As the Advanced Driver Assistance System (ADAS) gets installed in the vehicle, however, the factors affecting the driver's emotion are expanding beyond the appearance to the entire run. ADAS is a system that helps people drive more safely by notifying drivers and actively controlling the vehicles in advance of a dangerous situation through visual, auditory, and tactile (Vlacic *et al.*, 2001; Jones, 2002; Bishop, 2005). Since ADAS often intervene in the human operations directly or indirectly, such as interrupting or changing the driver's control actions (Strand *et al.*, 2014), driver's interaction with the system can induce negative feelings, which can then ruin the driver's experience. On such a background, this study aims to investigate the intrusive feelings caused by the Lane Keeping Assistance System (LKAS), which is a type of ADAS most popularly deployed recently by premium car brands. Intrusive feelings can be defined as unusual and undesirable feelings that can adversely affect the driver's experience as a result of unexpected actions of the ADAS. To minimize these negative feelings, it is necessary to calibrate parameters related to the behaviour of LKAS control. In this study, both expert designers, the designated personnel of the manufacturer, and normal drivers participated in the LKAS evaluation.

---

\* *Corresponding author*. e-mail: kjy@postech.ac.kr

In addition, suggested in this study is a metric as well as a statistical data analysis method to quantitatively measure the intrusive feelings through the vehicle sensor log data.

The paper is structured as follows. Section 2 describes the related work including LKAS concept. Section 3 then defines the research question and methodology. Section 4 demonstrates the results, and key implications and limitations are discussed in Section 5.

## 2. RELATED WORK

Lane Keeping Assistance System (LKAS), one of the ADAS, detects the lane marker with a front camera and keeps track of the relative position between the vehicle centre and the lane. Then, LKAS maintains the driving lane and reduces the risk of a lateral collision accident by means of automatic steering angle control using motor-driven power steering (MDPS) when unintended lane departure is expected (Risack *et al.*, 2000; Rajamani, 2012; Marino *et al.*, 2012).

To date, the evaluation of LKAS has been based mostly on functional completeness such as lane recognition (Mineta *et al.*, 2003) and lane departure rate (Hwang *et al.*, 2008). In ISO 11270 (2013), which contains performance test procedures, determines success if the outside of the tire does not deviate by more than offset (passenger car: 0.4 m, bus and truck: 1.1 m) from the lane boundary. On the other hand, there are evaluation guidelines that use lateral speed as a performance indicator. The maximum lateral velocity that does not deviate 0.5 m from the boundary of the outer lane is used as the performance index while raising the lateral velocity 0.5 m/s in the straight section (National Highway Traffic Safety Administration, 2013).

There are also several studies that demonstrate the effectiveness of LKAS such as human workload mitigation. Tanaka *et al.* (2000) reported that LKAS and Adaptive Cruise Control (ACC) were effective in reducing driver workload by measuring vital reaction. Blaschke *et al.* (2009) argued that there be a benefit to the lane keeping assistance function is situations that cause the distraction during actual driving such as phone dialling task.

The customer's expectation is now heading towards a holistic experience beyond the functional completeness and effectiveness of the technology. Even if LKAS is a safety-assisting function, providing the driver with an unfavourable experience may result in the driver turning the function off, which may ultimately render the supporting function useless. Therefore, it is very important to design the assistance system considering drive experience, but there is little research on experience and emotion of LKAS. Eichelberger and McCartt (2016) collected experience on ACC, Front Car Warning (FCW), Lane Departure Warning (LDW), and Lane Keeping Assistance (LKA), which are representative ADAS, through telephone interviews from drivers. They pointed out that the most annoying thing is the LKA, but there was no further analysis that why they regard it as an annoy, and there was a limit only depending on the driver's memory. This study focuses on the intrusive feelings of LKAS to identify factors that adversely influence the driver's affect.

## 3. METHODOLOGY

This paper focuses on uncovering the intrusive feelings affecting driver's affect and measuring them objectively.

### 3.1. Research Questions
We aim to answer the following research question:

- RQ1: What are the types of the intrusive feelings caused by LKAS?

As mentioned in Section 2, since continuous steering torque generated by LKAS intervenes the steering wheel, different vehicle behaviour and handle movements occur compared to the situation without LKAS. It can thus have an influence on drivers' affect as the result of perception with intrusiveness. For example, when the direction of handle operation is different between the driver and LKAS, the driver may feel discomfort due to interference. Therefore, this study investigates the intrusive feelings caused by LKAS.

- RQ2: Is it possible to derive variables, measuring the intrusive feelings objectively, from vehicle log data?

In order to measure the performance of the vehicle, subjective evaluation of experts was used in the development and production process. However, as vehicle systems become more intricate, the complexity of assessment is also increasing. Also, more time is required to perform the run test. These reasons are making the subjective evaluation harder and harder. In this study, we focus on vehicle log data for objective measurement of the intrusive feeling. As the development of electronics technologies in the automotive field, many different sensors including yaw rate sensor, torque measuring sensor on steering are used to measure the condition and state simultaneously in several modules at a cycle of 100

Hz. We try to derive variables that can express the intrusive feelings in the so-called big sensor data.

- RQ3: Is it possible to quantitatively compare the measured feelings?

The performance of LKAS is affected by performance of hardware components such as camera and lane recognition algorithm, but it is closely related to the parameter value (such as increase of intervention torque, lane departure distance performing torque intervention) controlled by LKAS designer. The ultimate goal of this study, minimisation of the intrusive feelings, can be used as a key strategy to optimise these parameter values in the context of LKAS designer. To do this, a quantitative comparison of intrusive feelings between LKASs set with different parameters must be solved first.

### 3.2. Research Flow

For solving the above-mentioned research question, there are three steps involved in the overall flow of the research: (1) Define, (2) Derive, and (3) Compare. The first step is to identify the intrusive feelings that can influence the driver's affect and define them as performance indicators. In the Derive phase, we select the signals that are closely related to the type of the intrusive feelings and derive as the new variables. Finally, we calculate the similarity score that can quantitatively compare the intrusive feeling between two LKASs.

#### 3.2.1. Step 1: Identifying and defining the intrusive feelings

A total of 17 drivers, including ten LKAS performance evaluation experts and seven normal drivers, participated a driving performance test session. Five vehicle models were used for the test, and about 6 hours of log data were collected. The total amount of raw data was about 18.9 GB (gigabytes) in CSV (comma-separated values) format. Various sensory interventions through visual, tactile, and vestibular senses were observed while test driving for approximately two hours on a highway with different LKAS parameters. The driving tests are conducted in two different conditions where driver puts their hand on the handle (grip condition) and does not put (non-grip condition). In the grip condition, we tried to observe the intrusive feelings when there exists mixture of LKA torque and driver's torque. In the other condition, we tried to detect the feelings when only LKA torque influences on the handle. To observe the internal and external environments well, a 360-degrees camera is used. The driver's vision is recorded with glass-typed eye-tracker with head mount camera.

After the driving test, an expert interview was conducted with human factor experts. The different types of intrusive feelings that can be sensed by visual, tactile, and vestibular senses were identified. Those were categorized as performance indicators.

#### 3.2.2. Step 2: Deriving the variables

In this step, variables that can describe the intrusive feeling are derived from the vehicle log data. The number of signals receiving from CAN (Controller Area Network) bus of the test vehicle is approximately 130 including many signals not related to LKAS such as headlamp, wiper operation, emergency light operation. Except for the signals, we selected only the ones that express the meaning of performance indicators and derived as new variables by manipulating one or more signal variables together for data analysis. The derived variables were examined through exploratory data analysis to see if there was a pattern difference between a good setting with little intrusive feeling and a bad setting.

#### 3.2.3. Step 3: Comparing the intrusive feeling using similarity

The empirical probability density function (PDF) was used to quantitatively compare the level of the intrusive feeling from data collected from different parameter settings in LKAS. PDF is widely used as a method of expressing temporal fluctuation patterns through empirically collected data (Freedman and Diaconis, 1981; Hidalgo et al., 2002). A histogram is first needed to be calculated to find the empirical PDF. In the histogram, it must define the bin size to count the number of observations in a specific range. To find the optimal bin size, there exist several methods including Sturges' formula, Rice rule, Doane's formula, Freedman-Diaconis' choice. However, in this research, the bin size is derived using heuristic methods of expert experiences. Lastly, the empirical PDF can be obtained by dividing the frequency of each bin by the total observation number $n$.

Finding the similarity between the two PDF there exist many different similarity measures (Cha, 2007) and we select intersection method (Duda et al., 2001), which is widely used a form of similarity. The similarity $s$ of PDF P and Q is defined as follows:

$$s = 100 \times \sum_{i=1}^{d} min(P_i, Q_i) \qquad (1)$$

where $d$ is the number of bins in the histogram. The similarity is expressed between 0 to 100, where 0 represents when the two histograms are different and 100 when the two are same. The intersection area of Figure 1 represents similarity. The performance indicator's score $p$ can be calculated by the weighted arithmetic mean of similarity obtained for each curve section (straight, low curve, and high curve):

$$p = \frac{\sum_{i \in \{str, low, high\}} w_i s_i}{\sum_{i \in \{str, low, high\}} w_i} \quad (2)$$

where $w$ is the weight, $s$ is the similarity. In this study, weights were given the same.

In order to verify the validity of the similarity-based comparison method, we examined the correlation between the subjective rating of drivers and the performance indicator's score based on empirical PDF.

## 4. RESULTS

### 4.1. Step 1: Identifying and Defining the Intrusive Feelings

#### 4.1.1. Visual perception
In curve section, a vehicle was not able to maintain its lane when controlling the handle with weak LKA torque. Drivers initially perceive the malfunction of the system with their eyes. Despite the LKAS being on, leaving the lane will give the driver a sense of disappointment and distrust of the system.

As one of the typical behaviours caused by LKAS, it is observed that the heading angle and steering angle of vehicles suddenly change due to the big torque on the steering wheel. Also, to recovery from the sudden changes, a vehicle tends to move in a zigzag pattern. Drivers were able to recognize that the vehicle moves unusually different with visual detection. Occasionally, the LKA torque intervenes at frequent intervals of more than about 1 Hz, causing the steering wheel to move like trembling. When drivers were not holding the handle, they could visually find it.

#### 4.1.2. Vestibular perception
The sudden change of steering angle comes together with the change of lateral speed, which leans to one side and drivers were able to feel this via their sense of equilibrium. This phenomenon can increase the anxiety of drivers.

#### 4.1.3. Tactile perception
When the trembling of steering wheel occurs, if drivers hold the handle, we observed they feel a vibration through the hand. The movement of the handle can be clearly distinguished from the natural handle movement or tremor caused by the road surface.

Also, when the direction of the torque that drivers provide is different from the LKAS, then drivers tend to feel that handles are rather heavy and are interfered by the torque to drive the vehicle normally. This interference torque seems to increase the driver's discomfort and fatigue.

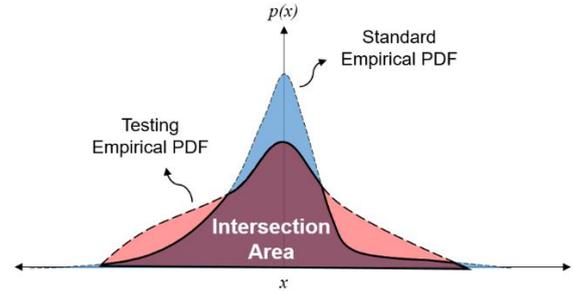

Figure 1. The example of similarity between two different empirical PDF.

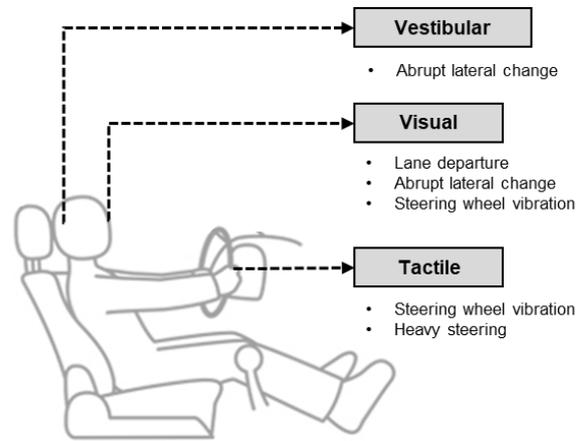

Figure 2. Types of the intrusive feelings caused by LKAS.

#### 4.1.4. Performance indicator
The intrusive feelings that felt from driver's various sensing organs (Figure 2) could be summarized in four categories including lane departure, abrupt lateral change, a vibration of the steering wheel, and heavy steering. The performance indicators that minimise those feelings can be defined as below:

- Lane keeping performance – Ability of the vehicle to maintain the centre of the lane without leaving the lane
- Dynamics behaviour stability – Ability of the vehicle to drive steady without sudden leaning to left or right
- Steering stability – Ability to minimize the vibration of steering due to the high-frequency intervention of LKAS on handles
- Non-interference performance – Ability to minimize the torque generated by the LKAS in the opposite direction to the driver's intention

## 4.2. Step 2: Deriving the Variables

### 4.2.1. Preprocessing

To make the raw data available for analysis, preprocessing was performed as shown in Figure 3. The raw data has the characteristic that the sensors record data independently. It means not every sensor recorded value simultaneously. Therefore, at a certain period, just a few of the signal data are recorded at that point; most of the other values of signal data are not recorded. Because there are so many null values in the raw data (Figure 4), time-based data reduction was conducted to handle efficiently the data in DBMS. In addition, among all the signal data, the most frequently collected data showed 100 Hz record rate. Therefore, we merge the data every 0.01 second. We assume that the empty values can be replaced by the most recently recorded value. For example, in the right table in Figure 4, the value of variable 2 at 0.02 second has the value of 3, which is the value coming from 0.0091 seconds, the most recently recorded value.

Moreover, because the operation of LKAS is influenced by the physical curvature of roads, we separate the curve type into three different sections, straight, low curve, and high curve. For this, we used the radius of curvature as a threshold, and the criteria are as following:

- Straight section: the radius of curvature 5000 m or higher
- Low curve section: the minimum radius of curvature 1000 m or higher in a section of the radius of curvature 5000 m or lower.
- High curve section: the minimum radius of curvature 1000 m or lower in a section of the radius of curvature 5000 m or lower.

Also, we filtered data where data are not related to intrusive feelings. Data which collected with LKAS off or not working properly, and intentionally changing the lane due to road situation are eliminated from data analysis. Then, we removed outliers of the measured variable.

### 4.2.2. Derived variables to represent the intrusive feelings

To analyse the lane-keeping performance, it is necessary to see how far a vehicle is away from the centre of the lane. In order to figure out how a vehicle changes laterally in a lane, we derived the vehicle's lateral position in the lane using the signal data of $distance_{left}$ and $distance_{right}$, which indicates the distance from the left and right side of the lane to the centre of the vehicle, respectively (Eq. 3). When the vehicle is located exactly at the centre of the lane, the value of LP is 0. When the vehicle is located on the left side of the lane, then the value is negative; the value is positive when it is located on the right side.

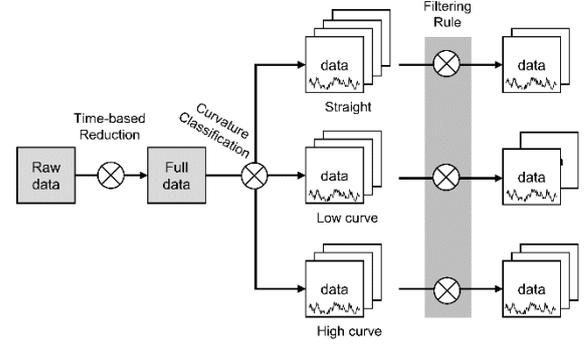

Figure 3. Overview of the vehicle log data preprocessing.

Figure 4. Time-based reduction in the vehicle log data preprocessing.

$$LP = \frac{distance_{left} - distance_{right}}{2} \quad (3)$$

Sudden lateral movement of a vehicle is caused by the lateral motion of vehicle, which can be experienced by the human vestibular organ. To see the vehicle's lateral direction, lateral speed was newly calculated using the vehicle's lateral position variable (LP), which is derived in the previous section. The lateral speed (LS) per unit time $t$ is defined as the following equation:

$$LS = \frac{\Delta LP}{\Delta t} \quad (4)$$

It is essential to observe the steering angle signal in order to capture the phenomenon of shaking steering wheel. To discover such pattern, the signal data of steering angle with high-pass band filter (Filtered Steering Angle, FSA) was derived as a new variable for analysis. We only passed frequencies higher than 1 Hz, which is the threshold frequency of steering wheel that drivers recognize as vibrating. We assumed that the frequencies lower than 1 Hz are caused by the normal steering control by LKAS.

As seen from the previous discovery, drivers experience the intrusive feeling when torque generated by LKAS is applied with the torques of opposite direction of the driver's. For example, when the driver rotates the steering wheel to the left side while the LKAS tries to rotate the wheel to the right side. To see this phenomenon on data, we derived a new variable called interference torque (IT) as follows:

$$IT = \begin{cases} l, \text{if } sign(l) \neq sign(d) \\ 0, \text{if } sign(l) = sign(d) \end{cases} \quad (5)$$

$$sign = \begin{cases} -1, \text{if } x < 0 \\ 0, \text{if } x = 0 \\ 1, \text{if } x > 0 \end{cases} \quad (6)$$

where $l$ is the amount of torque caused by LKAS and $d$ is the amount of torque caused by drivers. A negative value represents a torque that would accelerate the steering wheel in a clockwise direction.

4.2.3. Exploratory data analysis

In this section, we compared the characteristics and patterns of derived variables by analysing the log data. According to evaluators' response, a good setting where intrusive feelings are hardly felt and a bad setting that is felt much were selected for exploratory comparison. For analysing lane keeping performance, dynamics behaviour stability performance, and steering stability performance, data were collected without driver putting their hand on handles to see the patterns due to the pure LKA torque, trying to exclude external factors as much as possible. For analysing non-interference performance, data were collected while the driver performed a natural steering wheel operation in order to measure the interference torque. The measurement time of the selected log data is shown in Table 1, and the descriptive statistics of the derived variables in selected log data are shown in Table 2.

- Lane keeping performance

As shown in Table 2, the standard deviation of vehicle position in the good setting is much lower than the bad setting. As the result of Levene's test, the differences in variance are statistically significant ($p < .001$) in all conditions of different types of curve. Figure 5 represents the typical pattern which can be found in the run chart. Considering that the total width of the driving road is approximately 3.3 m, patterns of vehicle's position variable shows that the vehicle keeps the centre of the lane very well in good settings, but there is a high variation of pattern in bad setting.

Table 1. Measurement time of selected log data.

| Performance | Setting | Measurement time (mm:ss) | | | Total (mm:ss) |
|---|---|---|---|---|---|
| | | Straight | Low Curve | High Curve | |
| Lane-keeping | Good | 06:20 | 01:21 | 06:49 | 14:30 |
| | Bad | 06:26 | 02:03 | 03:45 | 12:14 |
| Dynamic Behaviour Stability | Good | 12:43 | 03:42 | 02:38 | 19:03 |
| | Bad | 06:31 | 02:03 | 02:40 | 11:14 |
| Steering Stability | Good | 06:26 | 02:03 | 03:45 | 12:14 |
| | Bad | 07:24 | 01:20 | 03:20 | 12:04 |
| Non-interference | Good | 21:18 | 08:07 | 09:29 | 38:54 |
| | Bad | 05:07 | 02:45 | 02:58 | 10:50 |

Table 2. Descriptive statistics of derived variables for each curve section.

| Variable | Setting | Straight | | Low Curve | | High Curve | |
|---|---|---|---|---|---|---|---|
| | | mean | s.d. | mean | s.d. | mean | s.d. |
| LP (m) | Good | -0.033 | 0.081 | -0.310 | 0.075 | -0.125 | 0.268 |
| | Bad | 0.356 | 0.554 | -0.289 | 0.580 | -0.019 | 0.765 |
| LS (m/s) | Good | -0.005 | 0.122 | -0.002 | 0.119 | 0.007 | 0.108 |
| | Bad | 0.009 | 0.275 | -0.022 | 0.335 | -0.025 | 0.329 |
| FSA (deg) | Good | $6.9 \times 10^{-6}$ | 0.001 | $-3.9 \times 10^{-6}$ | 0.001 | $1.9 \times 10^{-5}$ | 0.001 |
| | Bad | 0.001 | 0.213 | -0.003 | 0.258 | 0.0005 | 0.304 |
| IT (Nm) | Good | -0.003 | 0.007 | $-9.8 \times 10^{-6}$ | 0.006 | $5.3 \times 10^{-6}$ | 0.005 |
| | Bad | -0.053 | 0.564 | -0.164 | 0.638 | 0.234 | 0.842 |

- Dynamics behaviour stability

For lateral speed, the standard deviation in the bad setting is higher than in the good settings in all cases of curve types, and the difference of the variance is statistically significant through Levene's test ($p < .001$). However, the standard deviation does not necessarily increase as the curvature of lane increases to the high curved lane.

- Steering stability

The standard deviation of filtered steering angle (FSA) in good settings is considerably lower than bad settings (Table 2), and there was a significant difference in the variance in all curve conditions ($p < .001$). A frequency analysis technique is used to detect the vibrating phenomena of the steering wheel. The most frequently used technique to analyse the discrete signals is Fourier transform, which gives a meaningful result when the defect is a stationary signal in the whole measurement time. However, the vibrating signal of steering angle does not occur throughout the whole time. We consider steering angle signal data as the non-stationary signal because the defect signal only happens in a certain period. Therefore, we use Short-Time Fourier Transform (STFT) to analyse the defect frequency of steering angle. STFT applies window function on the signal that is to be

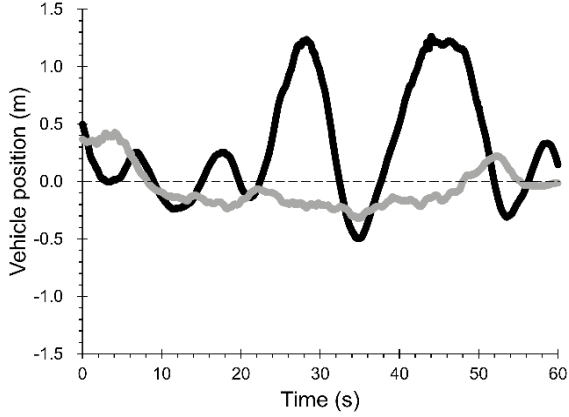

Figure 5. An example of run chart of vehicle position (grey: good setting, black: bad setting).

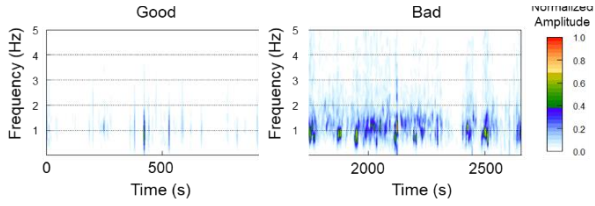

Figure 6. Spectrogram of Filtered Steering Angle signal.

analysed and performs Fourier transform. STFT can be defined as the following equation:

$$STFT\{x(t)\}(\tau, \omega) \equiv X(\tau, \omega)$$
$$= \int_{-\infty}^{\infty} x(t)w(t-\tau)e^{-j\omega t} dt \quad (7)$$

where w(t) is the window function, we used a Hann window, and ω is the frequency axis, τ is the time axis. x(t) is the input signal to be transformed. The result of STFT, which is the magnitude amplitude, is visualized using spectrogram with the following equation:

$$spectrogram\{x(t)\}(\tau, \omega) \equiv |X(\tau, \omega)| \quad (8)$$

The result of applying STFT on FSA signal is illustrated in Figure 6. Higher amplitude is observed in bad settings than good settings. For amplitudes above a certain level, we believe that drivers are able to feel the tremor of the steering wheel of the corresponding frequency domain.

- Non-interference performance

The statistics in Table 2 report that standard deviation of interference torque in good settings is smaller than the torque in bad settings, in all cases of curve types. Similarly, the differences of variance show the statistically significant level using Levene's test (p < .001).

Table 3. Measurement time of vehicle log data on various setting for validation of the similarity-based method.

| Condition | Setting | Measurement time (mm:ss) | | | Total (mm:ss) |
|---|---|---|---|---|---|
| | | Straight | Low Curve | High Curve | |
| Without grip the handle | A | 16:15 | 03:38 | 08:16 | 28:09 |
| | B | 06:26 | 02:03 | 03:45 | 12:14 |
| | C | 06:31 | 02:03 | 02:40 | 11:14 |
| | D | 01:42 | 00:00 | 03:54 | 05:36 |
| | E | 05:37 | 02:31 | 03:21 | 11:29 |
| | F | 04:05 | 02:52 | 00:52 | 07:49 |
| | G | 09:17 | 01:49 | 03:54 | 15:00 |
| | H | 13:46 | 02:30 | 04:33 | 20:49 |
| With grip the handle | I | 04:00 | 04:34 | 00:00 | 08:34 |
| | J | 01:55 | 03:46 | 00:00 | 05:41 |
| | K | 01:06 | 00:12 | 00:27 | 01:45 |

4.3. Step 3: Comparing the Intrusive Feelings using Similarity

In order to compare the score obtained using empirical PDF with the subjective rating of the drivers, the log data was collected from 11 runs carried out on the highway through vehicle model C. We collected about 112 minutes of data through 8 runs without handle grip condition, and about 16 minutes of data through 3 runs with handle grip condition. From the aspect of the curve type, we collected about 71 minutes of data in the straight section, about 26 minutes of data in the low curve section, and about 32 minutes in the high curve section. The total amount of raw data was about 6.57 GB in CSV format, and the amount of preprocessed data was about 0.31 GB in CSV format. The detailed collection time of log data is shown in Table 3. The subjective ratings were obtained by the average score of satisfaction in continuous scale between 0 and 100.

A scatter plot to illustrate the relationship between the subjective rating and the quantitative score is shown in Figure 7. As a result of correlation analysis, similarity score showed a tendency similar to subjective rating from drivers in dynamic behaviour stability ($r_p = 0.877, p < .01$) and steering stability ($r_p = 0.922, p < .01$). In the case of lane-keeping performance, although a weak correlation was revealed, linearity was not statistically significant ($r_p = 0.619, p = 0.10$). In the case of non-interference performance with grip handle condition, the sample number was insufficient to judge the significance, but the Pearson correlation coefficient r was high ($r_p = 0.944, p = 0.21$).

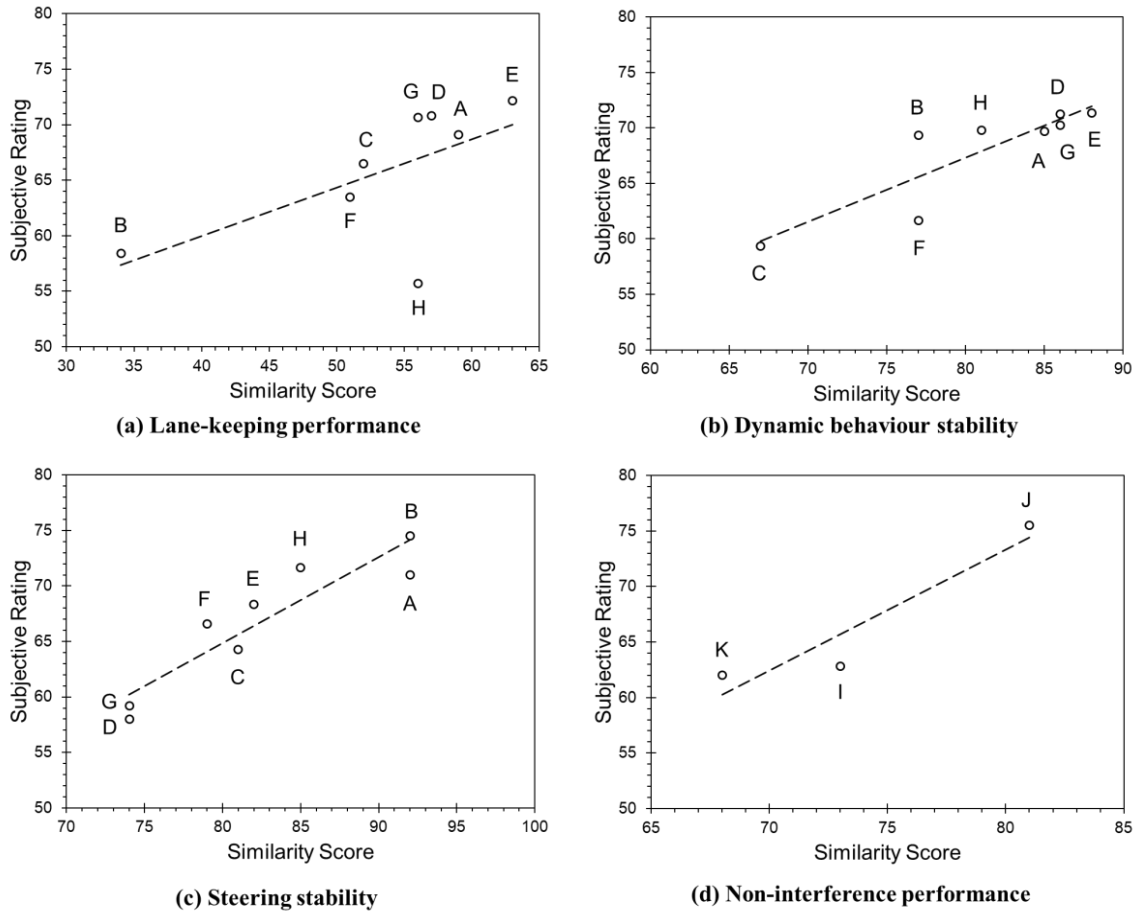

Figure 7. Scatterplot of subjective rating (Y) against quantitative score (X). The regression line is plotted in dotted line: (a) intercept = 42.61 (p < .05) and slope = 0.435 (p = 0.10), (b) intercept = 20.97 (p = 0.093) and slope = 0.579 (p < .01), (c) intercept = 2.802 (p = 0.81) and slope = 0.776 (p < .01), (d) intercept = -13.82 (p = 0.71) and slope = 1.058 (p = 0.21).

## 5. DISCUSSION

This study revealed the unpleasant, intrusive feeling that occurs when LKAS intervenes in driver 's driving task. The type of intrusive feelings was suggested in the form of performance indicators, allowing the LKAS to be evaluated on a variety of criteria that take into account the driver's affect, beyond only lane-keeping performance. In addition, a method for quantitative comparison based on objectively measured variable was suggested. The method can contribute to finding the optimal parameter of LKAS. We discovered that there was a trade-off among the proposed four performance indicators in controlling the values of LKAS parameters. For example, if we increase the amount of LKA torque, the vehicle can firmly keep the centre of the lane. However, the stability of vehicle becomes worse, and the shaking of steering wheel appears. Conversely, reducing the amount of LKAS torque resulted in lower lateral speed, improved behavioural stability and tremor, but failed to maintain the centre of the lane or deviated from the lane. LKAS designers who participated in the subjective rating also answered that the process of searching for parameters by considering these trade-offs would be a time-consuming and challenging task. It is expected that LKAS designers can employ our quantitative method to help find the optimal LKAS parameter.

It should be noted that the method proposed in step 3 is a relative comparison methodology, not an evaluation of the absolute performance of the system. Performance is represented by the similarity score obtained from the distribution of data. Therefore, standard data is essential to utilize this analytical methodology, and reliability of standard data is also very important. If the standard reference is not good enough and the better-performing data comes in, the quantitative score may be rather low. In this case, it is expected that we can find data having better performance than standard if we employ the auxiliary indicators such as kurtosis and kurtosis.

In setting H of Figure 7(a), meanwhile, there is a large difference between the regression line and the evaluation result, which seems to be due to lane departure. Although only one lane departure event occurred in the setting H, LKAS maintained the lane well in general. It seems that the evaluators had rated the lane-keeping performance significantly lower, as they are lethal to lane departure events. The reanalysis of the remaining 7 settings, where no lane departure was observed, showed a tendency between the two scores ($r_p$ = 0.945, $p < .01$). To compensate for this phenomenon, the weighting method reflecting the heuristics of experts can be considered. For example, it gives a larger weight to the difference of PDFs at a big lane displacement. However, in order to determine the weight, it is necessary to perform additional follow-up studies to identify the model between the lane departure distance and the perceived performance of the drivers.

## 6. CONCLUSION & FUTURE WORK

Designing a system that takes into account user affects is now a crucial issue in the automotive industry. In this paper, we describe the types of intrusive feelings that affect user emotion and build performance indicators to design driver-centric LKAS.

In addition, exploratory data analysis techniques were applied to real vehicle log data to reveal various patterns and features of intrusive feelings. Based on this, we proposed a similarity-based quantitative comparison methodology to help minimise the intrusive feelings. Our method is expected to complement the existing vehicle performance evaluation based on traditional survey method. Moreover, our method can support designers' efficient for optimal performance because they can see how much performance has improved as they adjust the LKAS parameters.

Our future research will focus on the development of analytics support software that can automatically perform the proposed quantitative analysis. Because the vehicle log data is big and massive, automated analysis and evaluation will be highly utilized in the industry. In the near future, various types of intrusive feeling are expected to be found that inhibit the driver 's affect in the automobile industry due to the development of autonomous driving technology. Research to identify this and create a better driving experience using massive log data should be conducted in both academia and industry.

**ACKNOWLEDGEMENT**−This study has been supported by Hyundai NGV and Hyundai Motor Company.